# Understanding the Effect of Uniaxial Tensile Strain on the Early Stages of Sensitization in AISI 304 Austenitic Stainless Steel

P. S. Chowdhury[a*], S. K. Guchhait[a], P. K. Mitra[a], P. Mukherjee[b], N. Gayathri[b] and M. K. Mitra[a]

[a]DEPARTMENT OF METALLURGICAL AND MATERIAL ENGINEERING,

JADAVPUR UNIVERSITY, KOLKATA -700032

[b]VARIABLE ENERGY CYCLOTRON CENTRE (VECC),

1/AF BIDHAN NAGAR, KOLKATA-700064

**Abstract:**

In the present study, an attempt has been made to understand the effect of different competing mechanisms controlling the overall degree of sensitization (DOS) of deformed austenitic stainless steel at the early stage of sensitization. The Double Loop Electrochemical Potentiokinetic Reactivation (DL-EPR) studies were performed to characterize the Degree of Sensitization (DOS) as functions of pre-defined strain and sensitization temperature. X-ray Diffraction (XRD) and Scanning Electron Microscopy (SEM) were used to explain the phenomena qualitatively. A non monotonous behaviour in the variation of DOS has been observed with deformation and sensitization temperature. The presence of Deformation Induced Martensites (DIM) and their transformation into tempered martensites ($\alpha$ + $Fe_3C$) at higher temperatures was found to play major roles in controlling the overall sensitization and desensitization processes.

[*]Corresponding author:

Dr. Prasun Sharma Chowdhury

Dr. D. S. Kothari Post Doctoral Fellow

Department of Metallurgical and Material Engineering

Jadavpur University, Kolkata – 700032

Email: psc0126@gmail.com

Phone: +91 980 451 8402

# 1. Introduction:

Austenitic stainless steels are widely used for different industrial purposes due to a combination of good mechanical properties and excellent corrosion resistance over a large temperature range [1-3]. The high chromium content in these alloys plays a vital role in achieving remarkable resistant properties against different types of uniform corrosions. However, there are also different types of localised corrosions which are observed in austenitic stainless steels [1-4]. Sensitization in austenitic stainless steels affects the uniformity of the chromium content throughout the microstructure by formation of chromium carbide precipitates along the grain boundaries, resulting in the depletion of Cr at near grain boundary regions. Intergranular corrosion is a type of localised corrosion frequently observed in different types of austenitic stainless steels, solely due to the process of sensitization.

Since sensitization is a diffusion assisted process, it is effectively influenced by the microstructure which is the result of different thermo mechanical processes experienced by the materials. In the work performed by L. E. Murr et al. [5], it was observed that the uniaxial tensile strain prior to aging of solution annealed AISI 316 stainless steel pipes resulted in a dramatic change in the degree of sensitization (DOS) of those materials. It was first observed by Povich and Rao [6] that AISI 304 stainless steel pipings in boiling water reactors experience significant sensitization after 10 years even at the operating temperature of 573K. Trillo et al. [7] have shown that not only the microstructure has significant influence on precipitation and sensitization kinetics, but also the deformation particularly the amount of straining may accelerate these kinetics processes [5,8-9]. The influence of α′- martensite (formed during deformation) and γ austenite serves as nucleation site for carbide precipitation and provide an effective grain refinement which accelerate both sensitization and desensitization [10]. However, in most of the recent works [11,12], the phenomenon of low temperature sensitization have been studied at relatively higher temperature (> 773K) and with longer ageing time. In the present study, the main objective is to understand the early sensitization process of 304 grade austenitic stainless steel due to prior straining of the as received as well as heat treated samples at a temperature regime of 473 – 873 K with soaking time of 1 hour. The microstructure of the samples has been characterised by Scanning electron microscope (SEM) and X-ray diffraction technique (XRD). The degree of sensitization has been measured by double loop electrochemical potentiokinetic reactivation (DL-EPR) test.

## 2. Experimental:

The AISI 304 stainless steel was obtained in the form of rolled sheet of 3 mm thickness. Tensile samples of gauge length 30 mm were machined and these were deformed at different uniaxial strains (10%, 20% , 30%, 40%, 50%) using the Instron$^{TM}$ tensile testing machine at a strain rate of $10^{-3}$ per second . The gauge length portion of each sample was then cut into four rectangular pieces and three of them were heat treated at temperature 473K, 673K and 873K. The heat treatment was performed under vacuum (air pressure ~ $10^{-3}$ mbar) and the soaking time was maintained at 1 hour followed by furnace cooling. A set of 24 samples were thus prepared from 5 set of tensile samples.

The XRD patterns were obtained from a Rigaku Ultima III X-ray Diffractometer using Cu-K$_\alpha$ radiation in the 2θ range of 30º to 100º.

The DL-EPR studies were carried out on all the 24 samples to measure their DOS. The specimens were finely polished and then ultrasonically degreased in soap solution. Then the samples were washed thoroughly with distilled water and dried. DL-EPR test were conducted in a polarization cell containing 0.5M $H_2SO_4$+0.01M KSCN solution. The mounted sample were immersed in the solution and the open circuit potential (OCP) of the specimen were noted. Inside the cell, the polarization of the specimen with respect to a saturated calomel electrode (SCE) was maintained to -500 mV for 2 minutes to dissolve the air formed oxide film. Then the specimen was polarized anodically from -500 mV (SCE) to +300 mV (SCE) and then cathodically polarized from +300 mV (SCE) to OCP at a scan rate of 1.67 mV/s using Gamry 600TM model electrochemical interface. The electrode potential vs. Log current was then recorded with the help of the interface. From the variation of current density with electrode potential (the DL-EPR curve), the peak current during reactivation (Ir) and the peak current during activation (Ia) was noted and the percentage ratio of them (Ir/Ia × 100%) were measured as DOS.

The JEOL JSM-6360 Scanning Electron Microscope (SEM) was used to obtain the micrographs of all the samples. Before taking the micrographs, the samples were finely polished and then etched using 10% HNO3 + 30% HCl + 60% distilled water solution.

## 3. Results:

XRD pattern has been collected in as received and heat treated condition. Fig. 1a shows the XRD peaks of as received sample and deformed sample as a function of strain. Split in (111) γ peaks reveals the presence of martensite in a small quantity within the matrix. A change is observed at 20% strain where a single peak of (111) γ is only observed with the suppression of adjacent (110) α′ peak due to preferred orientation. The split in the peak is again observed at 30% strain and the intensity of this peak increases as a function of deformation. The similar phenomena were observed for the sample heat treated and aged at 473K (fig. 1b), where strain induced martensite formation is seen at a relatively lower strain (i.e. at 20%). The martensite is found to exist up to 50% strain without decomposition and its intensity increased with deformation due to preferred orientation. However at 673K the α′ martensites present in the sample start transforming into tempered martensite (α + $Fe_3C$) which results in the phase mixture of α and α′. As a result the intensity of XRD peaks corresponding to α′ phases substantially decreased which is evident from fig. 1c. The transformation becomes more predominant at higher temperature, resulting in the formation of substatial amount of α phases in γ austenite matrix at the temperature of 873K. Hence at that temperature a measurable increase in the intensity of the XRD peaks adjacent to the (111)γ peaks is observed in the fig. 1d due to the presence of α phase in the sample.

A typical DL-EPR curve for the highest strained as deformed sample is represented in fig. 2. From this figure, it is clear that the height of the reactivation peak ($I_r$) is significantly smaller than the height of the activation peak ($I_a$), resulting in a measurable value of DOS for this sample.

The variation of DOS with respect to percentage strain for all the deformed and heat treated samples along with the as received one is shown in fig. 3. No significant variation in the DOS was observed at the room temperature. However, at higher temperatures, the variation of DOS with strain becomes significant and changes with increasing temperature. A systematic increase in DOS with increasing strain is observed after 20% pre-strain, especially at the temperature of 873K. However, the nature of variation of DOS with strain is almost similar at temperatures 473K and 673 K.

Fig. 4 represents the variation of DOS with respect to temperature. It was observed that, though the DOS values at room temperature for all the samples are almost same, their nature of variation with temperature changes drastically depending upon their extent of deformation.

The Scanning Electron Microscopy (SEM) was done on all the samples to study the microstructural change with respect to strain and sensitization temperature. Some typical Scanning Electron Micrographs for the as received and also for the deformed samples are shown in fig. 5. A small amount of deformation bands is observed (fig 5) even in the as received samples. From fig. 5, significant change in the microstructure of the sample was observed with increasing degrees of deformation. However, there is an increase in the population of deformation bands which initiates significantly from the sample with 20% strain. Different scanning electron micrographs obtained from the selected samples after the heat treatment at different temperature is shown in fig. 6. The micrographs in fig. 6 reveal the presence of deformation bands at all the temperatures.

## 4. Discussion:

The process of sensitization and desensitization is primarily governed by the kinetics of the formation of chromium carbide precipitates along the grain boundaries [13-15]. However, recent researches revealed that the deformation bands formed inside the grains of austenitic stainless steels are also significantly responsible since they act as favourable regions for chromium carbide precipitation; thus resulting in new types of sensitization and desensitization processes which take place inside the grains [16]. Especially in case of AISI 304 austenitic stainless steels, these deformation bands act as significant sources for the nucleation of α′-martensites [16], resulting in an increase in the population of α′-martensites inside the grains. When sufficient thermal energy is supplied in addition (up to 773K), these highly populated strain induced martensites (SIM) recrystallize very fast and form very small phase mixtures of α′-martensites and γ-austenites. The interphase boundaries between these two phases then act as the nucleation sites for carbide precipitation and thus accelerates both sensitization and desensitization processes [10]. It was reported [17-20] that the Martensite Induced Sensitization (MIS) in cold deformed AISI 304 austenitic stainless steel can be initiated even by a short term exposure to temperatures from 523K – 773K.

Beside the above mentioned sensitization process, a new kinetics of sensitization is also possible inside the grains when the sample is heated above 673K. After this temperature, the α′-martensites present in the sample start transforming into tempered martensites (α + $Fe_3C$). As the temperature rises, more amount of α′-martensites gets transformed into α phase. Hence in this case, a fine α - γ (bcc – fcc) phase mixture is formed along with $Fe_3C$ and the interphase boundaries between them play a vital role in controlling the sensitization process inside the grains, even at higher temperatures. Moreover the classical sensitization process which is known to be dependent upon the carbide precipitation along grain boundaries is also observed after the heat treatment above 773K [21].

The diffusion of chromium atoms through the grain matrix is an important factor which effectively controls both sensitization and desensitization process. Though it is well observed that the increase in prior deformation accelerates the sensitization process [7,20,22-26], there are also reports [12,27] in support of the non monotonous behaviour between cold work and sensitization. In fact, besides the increase in the rate of sensitization, higher deformation also enhances the diffusion of chromium which in turn increases the desensitization process. This is due to the fact that higher deformation leads to generation of more dislocations and point defects in the microstructure which helps the chromium to diffuse more rapidly from the matrix towards the chromium depleted zone. Since the process of sensitization and desensitization are mutually opposite phenomena, the overall effect of them on the degree of sensitization makes it non monotonous with increasing deformation which is observed distinctly in fig.3 and fig. 4.

Hence the above phenomena can be summarized as follows. There are two distinct types of sensitization and desensitization processes which take place in AISI 304 austenitic stainless steels. The first type covers the classical sensitization and desensitization processes which depend upon the formation of chromium carbides along the grain boundaries and the diffusion of chromium atoms towards the chromium depleted near grain boundary regions. The second type is the sensitization and desensitization processes which occur inside the grains. The α′ and α induced sensitization and desensitization are the parts of the second type. However unlike the α′ induced sensitization and desensitization processes, the α induced processes can also be observed at higher temperatures. Hence at higher temperatures (above 773K), both types of processes are involved.

The presence of α′-martensite phase in as received AISI 304 austenitic stainless steel samples and its variation with degree of prior deformation and transformation with sensitization temperature is clearly revealed by the XRD patterns (fig. 1). The variation in DOS in fig. 3 and fig. 4 below the temperature of 673K can be attributed to martensite induced sensitization and desensitization process. However from 673K, the α induced sensitization and desensitization process takes place simultaneously. Fig. 3 and fig. 4 shows a highly non monotonous variation in DOS with increasing strain and temperature at the early stages of sensitization. Since sensitization and desensitization are two mutually opposite dynamical processes, the short time exposure (1hour) of the samples could not stabilize these processes and as a result, the variation in DOS at the early stages of sensitization has been found to be non monotonic. Moreover, a significant amount of α′-martensite was observed even in the as received sample (fig. 1a) which may be responsible behind their prior sensitization as observed in fig. 3 and fig. 4. However, the values of DOS have increased marginally with increasing deformation at room temperature due to the absence of sufficient thermal energy necessary to initiate the sensitization and desensitization processes.

From fig. 3, it is observed that the values of DOS at 473K and 673K are higher than those of as received samples in the range of 10% to 40% strain. This may be attributed due to the martensite induced sensitization process dominating over the desensitization process resulting in an increase in DOS for these samples. However, the lower values of DOS for the samples with highest deformation (50% strain) clearly reveal the dominance of the desensitization process over the martensite induced sensitization. It is also interesting to observe that at the temperature of 873 K, the values of DOS slowly decreases up to 20% strain and then follows a steep rise with increasing deformation. At this temperature, beside the α induced sensitization and desensitization processes, the DOS values are also controlled by the classical grain boundary sensitization and desensitization processes. Hence the observed phenomena could be explained as a result of four dynamical processes (i.e. sensitization and desensitization processes within grains and at grain boundaries having different microstructural variation as explained above) taking place simultaneously. In fig. 3, the dominance of the overall desensitization processes results in the slow decrease in DOS up to 20% strain at the temperature of 873K. However with further increase in deformation, the overall sensitization process (both α induced and classical sensitization) predominates over the overall desensitization process, resulting in a steep rise in DOS.

Fig. 4 clearly reveals the effect of strain on the nature of variation of DOS as a function of temperature. The DOS of the as received sample did not vary much with increasing temperature. However a little fall in DOS value at a temperature of 473K compared with the room temperature DOS value clearly indicates the effect of desensitization on the as received sample. With increasing degree of deformation, the population of deformation induced martensites (DIM) as well as the population of dislocation networks increases, which in turn enhances the chromium diffusion process, causing a change in the nature of variation of DOS with increasing temperature. From fig. 4, it is interesting to note that at the highest temperature (873K), the desensitization process predominates for the samples with lower deformation (10% and 20% strain). However, for the samples with higher deformation (30%, 40% and 50%) the sensitization process overrides the desensitization process, showing an increase in DOS in fig. 4.

Fig 5 shows the evidence of deformation bands which are found to increase with increasing deformation. These bands still exist with the increase in temperature as seen in the micrograph of fig. 6.

**5. Conclusion:**

1. The DOS of AISI 304 austenitic stainless steel samples obtained from DL-EPR studies were successfully characterized as functions of prior deformation and sensitization temperatures.

2. XRD studies on the samples revealed the presence of DIM which was found to be responsible in controlling the low temperature sensitization process.

3. The two mutually opposite dynamical processes i.e. martensite induced sensitization and desensitization were found to take place at the earliest stage which resulted in a non monotonous variation in DOS with increasing deformation and sensitization temperature up to 673K.

4. From 673K, the transformation of $\alpha'$ to $\alpha$ has also been found to be responsible for the sensitization and desensitization processes. The dynamics involved behind this new type of process ($\alpha$ induced) has been addressed in the present work which also plays a vital role for sensitization and desensitization at higher temperatures.

5. The SEM micrographs obtained from the deformed and heat treated samples showed the presence of deformation bands for all the samples.

**Acknowledgement:**

One of the authors (P. S. Chowdhury) wants to acknowledge UGC, India for providing financial support.

**List of figures:**

1. Fig. 1: XRD full profiles of AISI 304 stainless steel samples as functions of strain and temperatures.

2. Fig. 2: A typical DL-EPR curve for the 50% pre-strained as deformed sample.

3. Fig. 3: Variation of DOS with strain at different temperatures.

4. Fig. 4: Variation of DOS with temperature at different strains.

5. Fig. 5: Scanning electron micrographs of AISI 304 as received and deformed samples.

6. Fig. 6: Scanning electron micrographs of AISI 304 samples at different strains and temperatures

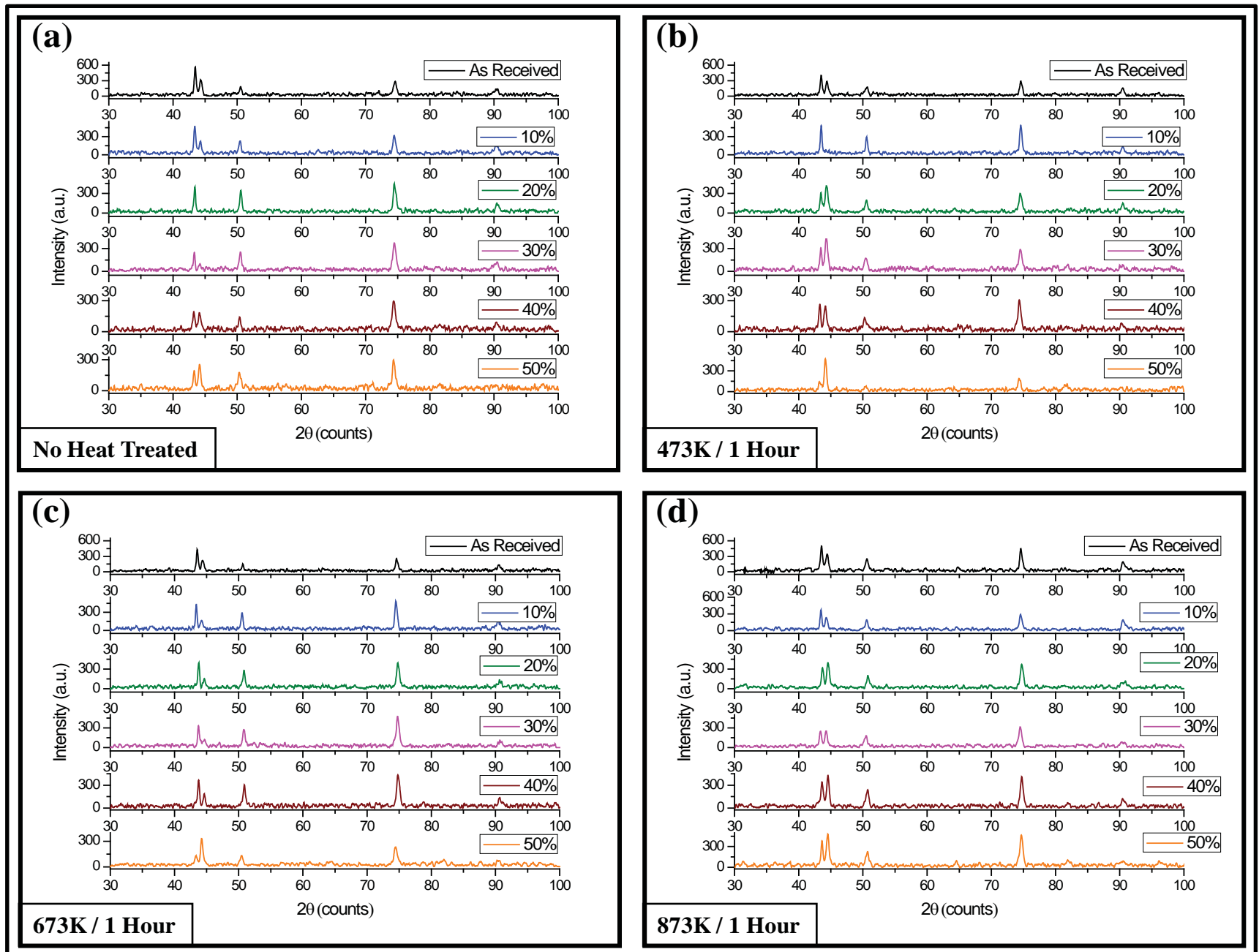

Fig. 1: XRD full profiles of AISI 304 stainless steel samples as functions of strain and temperatures

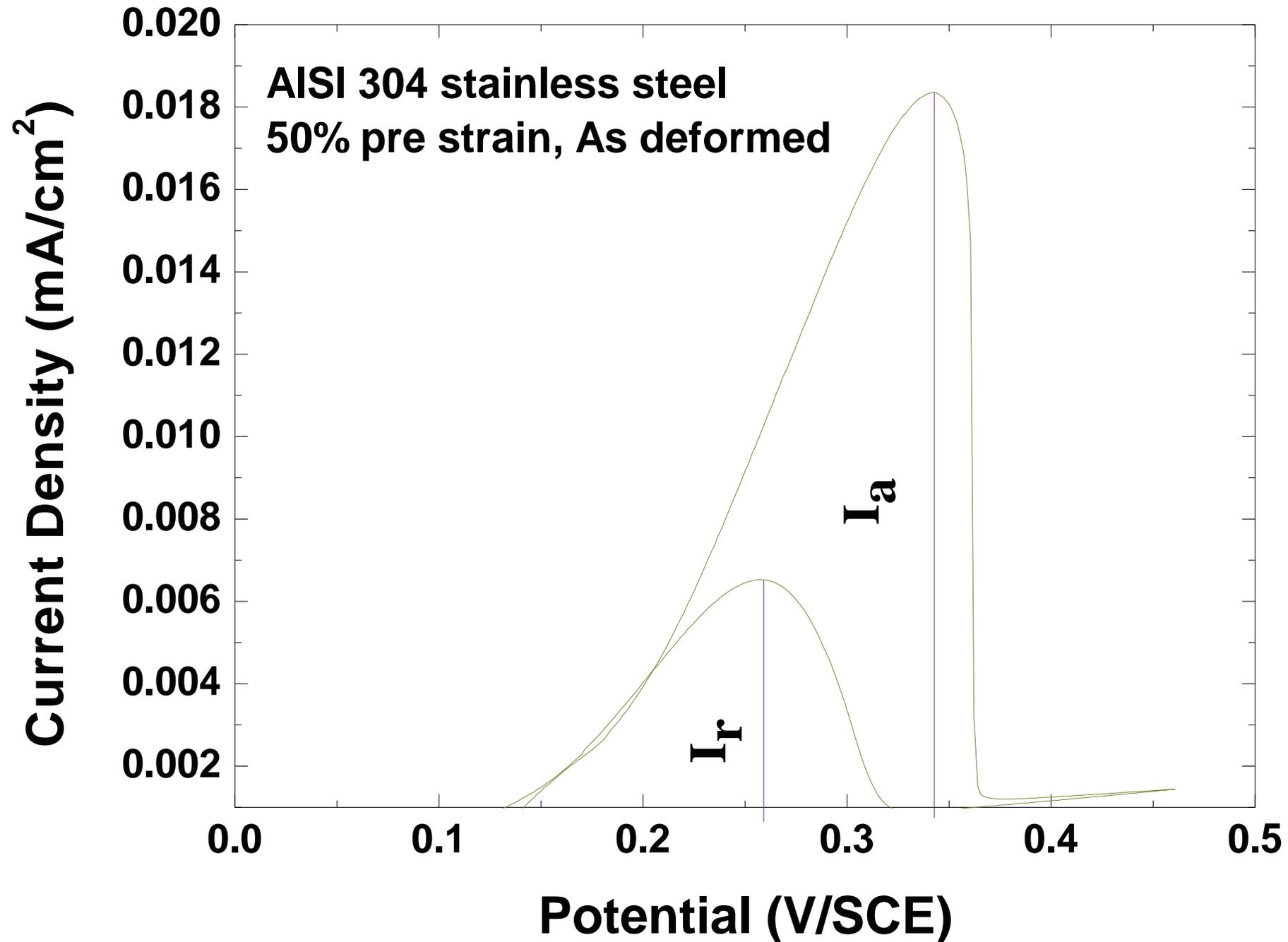

Fig. 2: A typical DL-EPR curve for the 50% pre-strained as deformed sample

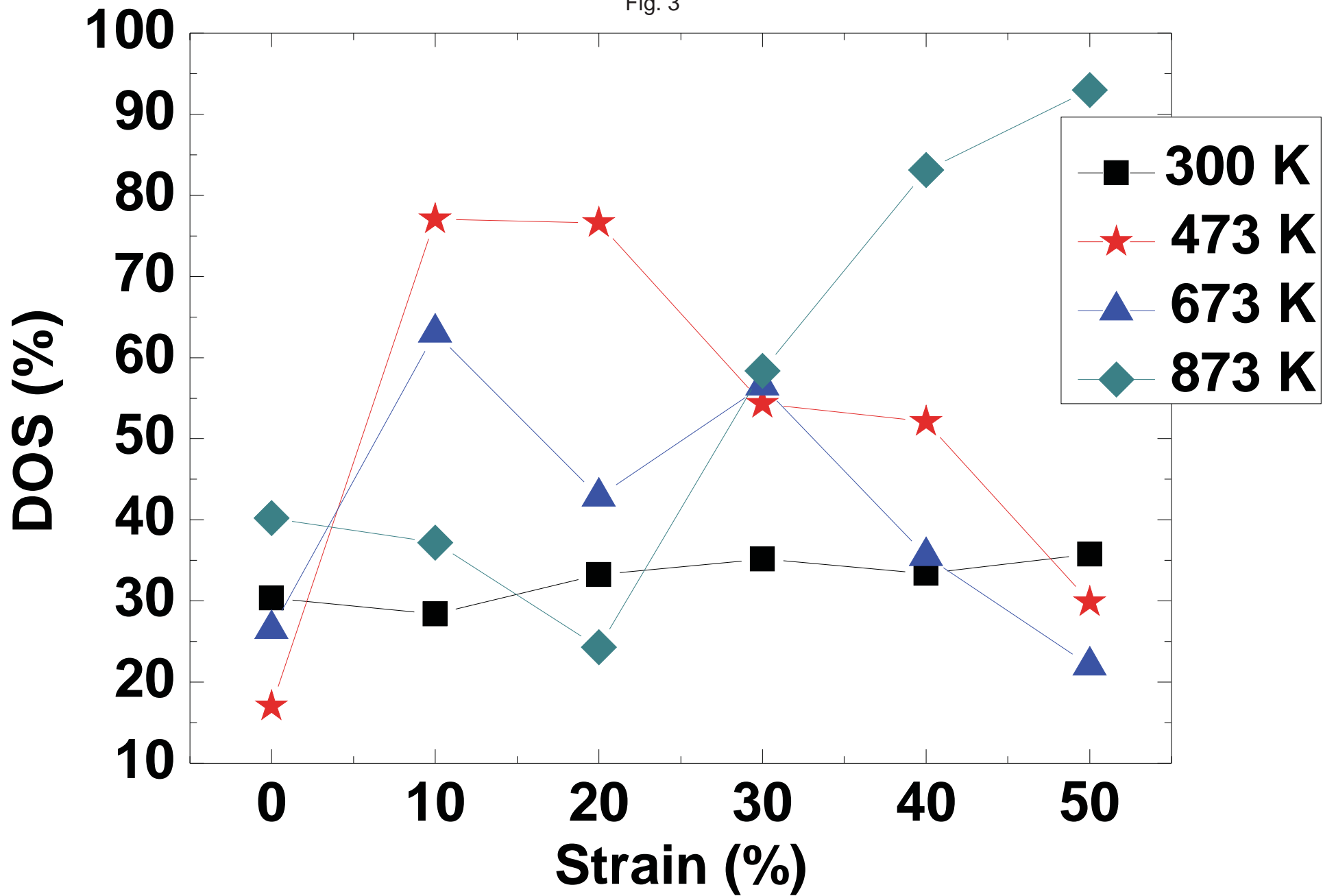

Fig. 3: Variation of DOS with strain at different temperatures

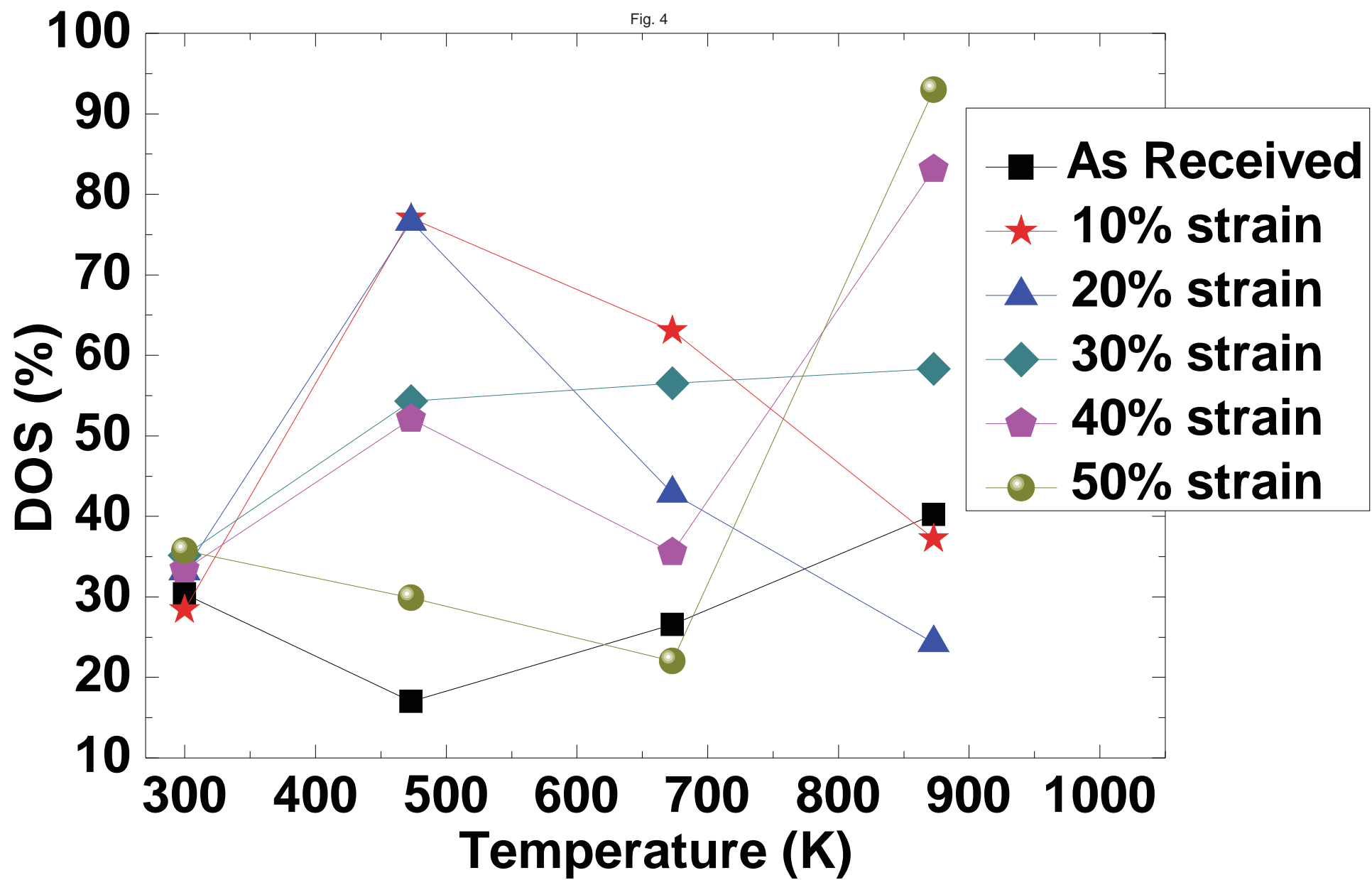

Fig. 4: Variation of DOS with temperature at different strains

Fig. 5

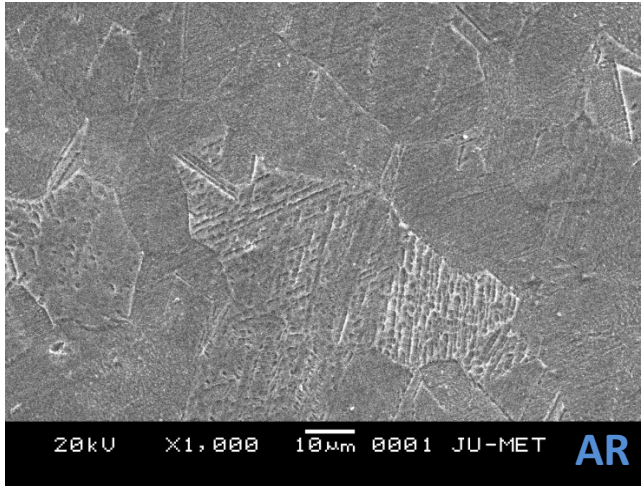
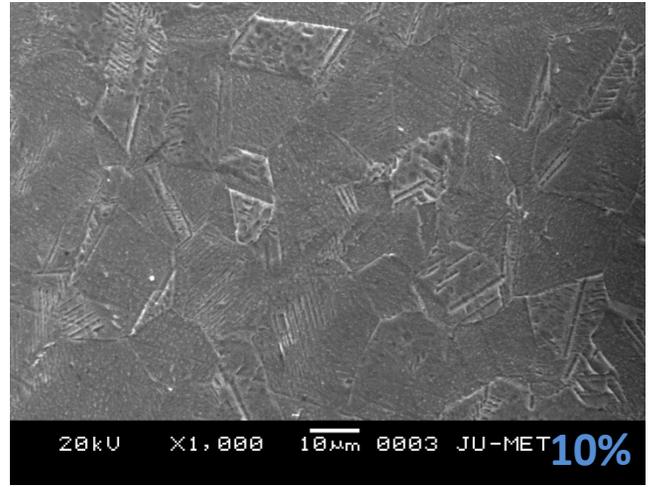
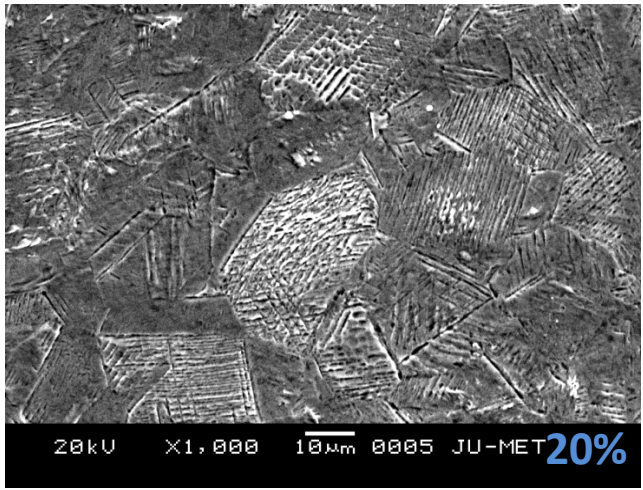
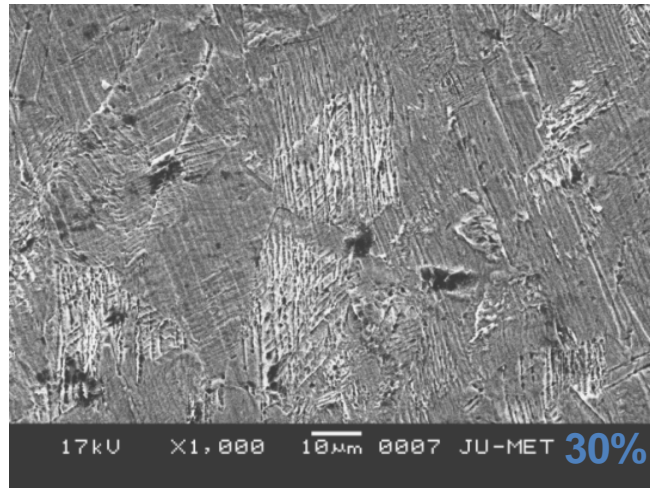
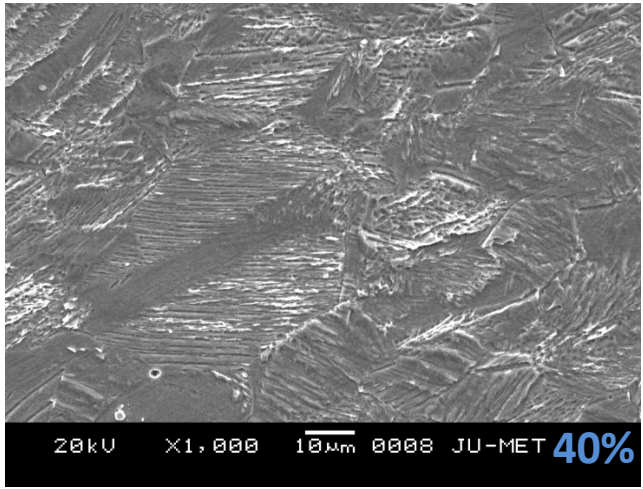
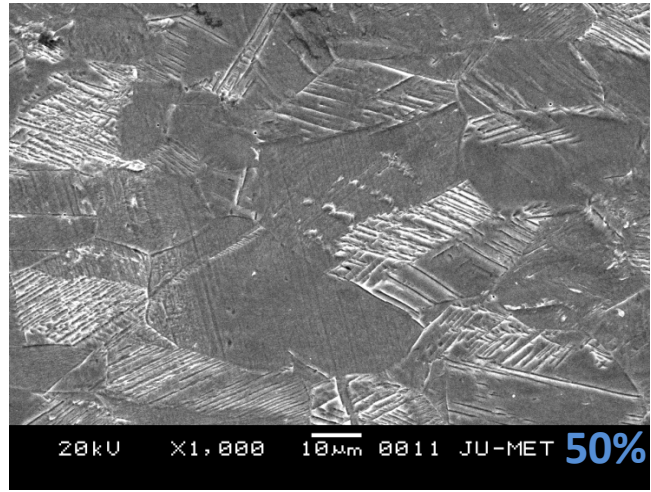

Fig. 5: Scanning electron micrographs of AISI 304 as received and deformed samples

Fig. 6

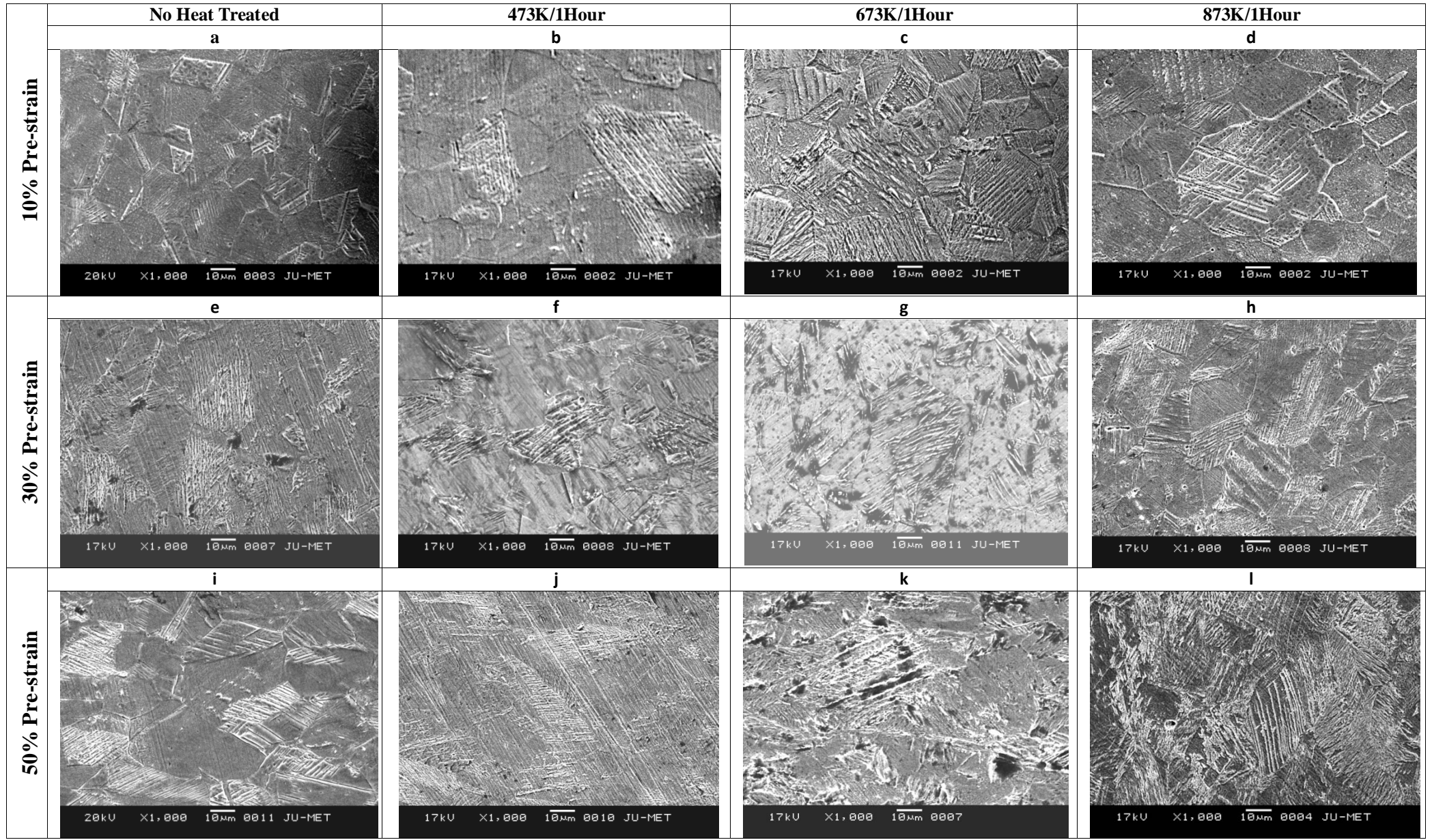

Fig. 6: Scanning electron micrographs of AISI 304 samples at different strains and temperatures